%% file: main.tex
\documentclass[review]{elsarticle}

\usepackage[]{algorithm2e}
\usepackage{lineno,hyperref}
\usepackage{listings}
\usepackage{amsmath}
\usepackage{multirow}
\modulolinenumbers[5]

\journal{Computers \& Security}

\begin{document}

\begin{frontmatter}

\title{Software Vulnerability Analysis Using CPE and CVE}


\author{Luis Alberto Benthin Sanguino}
\ead{lbenthins@gmail.com}
\author{Rafael Uetz}
\ead{rafael.uetz@fkie.fraunhofer.de}
\address{Fraunhofer FKIE, Bonn, Germany}



\begin{abstract}
\input{abstract}
\end{abstract}

\begin{keyword}
\texttt{elsarticle.cls}\sep \LaTeX\sep Elsevier \sep template
\end{keyword}

\end{frontmatter}


\section{Introduction}
\label{sec:introduction}
\input{introduction}

\section{Background}
\label{sec:background}
\input{background}	{

\section{CPE Dictionary and CVE Feeds Analysis}
\label{sec:analysis}
\input{analysis}

\section{Method}
\label{sec:method}
\input{method}

\section{Evaluation}
\label{sec:evaluation}
\input{evaluation}

\section{Related Work}
\label{sec:related_work}
\input{related_work}

\section{Conclusion}
\label{sec:conclusion}
\input{conclusion}

\bibliography{references}

\end{document}

%% file: abstract.tex
In this paper, we analyze the Common Platform Enumeration (CPE) dictionary and the Common Vulnerabilities and Exposures (CVE) feeds. These repositories are widely used in Vulnerability Management Systems (VMSs) to check for known vulnerabilities in software products. The analysis shows, among other issues, a lack of synchronization between both datasets that can lead to incorrect results output by VMSs relying on those datasets. To deal with these problems, we developed a method that recommends to a user a prioritized list of CPE identifiers for a given software product. The user can then assign (and, if necessary, adapt) the most suitable CPE identifier to the software so that regular (e.g., daily) checks can find known vulnerabilities for this software in the CVE feeds. Our evaluation of this method shows that this interaction is indeed necessary because a fully automated CPE assignment is prone to errors due to the CPE and CVE shortcomings. We implemented an open-source VMS that employs the proposed method and published it on GitHub.

%% file: introduction.tex
Cybercriminals often infect their victims by exploiting vulnerabilities in software products. For example, a cybercriminal can persuade a user to open a malicious PDF document using social engineering techniques like spear phishing. If the user employs a vulnerable version of a PDF reader, malicious code embedded in the document exploits the vulnerability, resulting in the user being infected with malware (e.g., Ransomware).

Vulnerabilities on software products seem to be a never-ending problem. In 2015, 5,585 new vulnerabilities were found \cite{symantec_16}. Similarly, in 2016, more than 6,000 software flaws were published in the National Vulnerability Database (NVD)\footnote{\url{https://nvd.nist.gov}}. The number of undisclosed or zero-day vulnerabilities has increased over the last years \cite{symantec_16}. Nonetheless, publicly known vulnerabilities still are the highest risk for organizations. As reported by the SANS institute, in more than 80\% of the security incidents known flaws were exploited \cite{sans_report_17}. The most exploited vulnerabilities in 2015 were vulnerabilities published in 2011 and 2007 \cite{verizon_dbir_16}.

To avoid data breaches caused by software vulnerabilities, organizations deploy Vulnerability Management Systems (VMSs) inside their IT infrastructures. Among the main functionalities, a VMS scans for software products installed inside a company that has potential vulnerabilities. To perform this task, the system manages an inventory containing information of the installed products, and moreover, it correlates this information with vulnerability information obtained from private (e.g., software vendors) and public sources, such as Computer Emergency Response Teams (CERTs), the NVD or Bugtraq\footnote{\url{http://bugtraq-team.com}}. 

Within the most used sources of vulnerability information are the Common Vulnerabilities and Exposures (CVE) feeds\footnote{\url{https://nvd.nist.gov/vuln/data-feeds}}, which are a subset of the NVD and offered by the National Institute of Standards and Technology (NIST). The CVE feeds contain identifiers for known vulnerabilities (e.g., CVE-2015-1538) and information related to them. To identify software products affected by a vulnerability, a CVE includes a list of vulnerable software, in which the description of a product follows the specifications of the Common Platform Enumeration (CPE) standard\footnote{\url{https://cpe.mitre.org}}.

The use of standards like CVE and CPE helps IT security vendors, security experts, research institutes, etc. to publish and exchange information efficiently and effectively. Additionally, the standards are employed by VMSs to assign structured identifiers to software products (inventory) and search for vulnerabilities related to them (vulnerability scanning). In this paper, we present an analysis of the official CPE dictionary and CVE feeds that show inconsistencies in these datasets that could prevent VMSs from delivering reliable results when scanning for vulnerabilities.

The paper is structured as follows. We first introduce the CPE and CVE standards and their role in VMSs (Section \ref{sec:background}). Then, we provide an analysis of the CPE dictionary and CVE feeds (Section \ref{sec:analysis}). In Section \ref{sec:method}, we describe a method that aims at dealing with the issues found in the CPE dictionary and CVE feeds. Section \ref{sec:evaluation} contains an evaluation of our method. Section \ref{sec:related_work} describes related work. Finally, Section \ref{sec:conclusion} gives a conclusion.

%% file: background.tex
This section briefly describes the standards that are used in our method to scan for vulnerabilities: Common Platform Enumeration (CPE) and  Common Vulnerabilities and Exposures (CVE). In addition, we explain the relation between both standards and how they are employed by VMSs.

\subsection{Common Platform Enumeration (CPE)}
CPE is a method that specifies a naming scheme for applications, hardware devices, and operating systems. CPE is part of the Security Content Automation Protocol (SCAP)\footnote{\url{https://scap.nist.gov}} standard, which was proposed by the National Institute of Standards and Technology (NIST). Currently, there exist two versions of the CPE specification: CPE 2.2 and CPE 2.3. Version 2.3 defines a stack formed by five specifications, including the CPE naming specification \cite{cpe_2_3_naming_specification} and the CPE dictionary specification \cite{cpe_2_3_dictionary_specification}. 

\begin{figure*}[!t]
\centering
\includegraphics[width=4.6in]{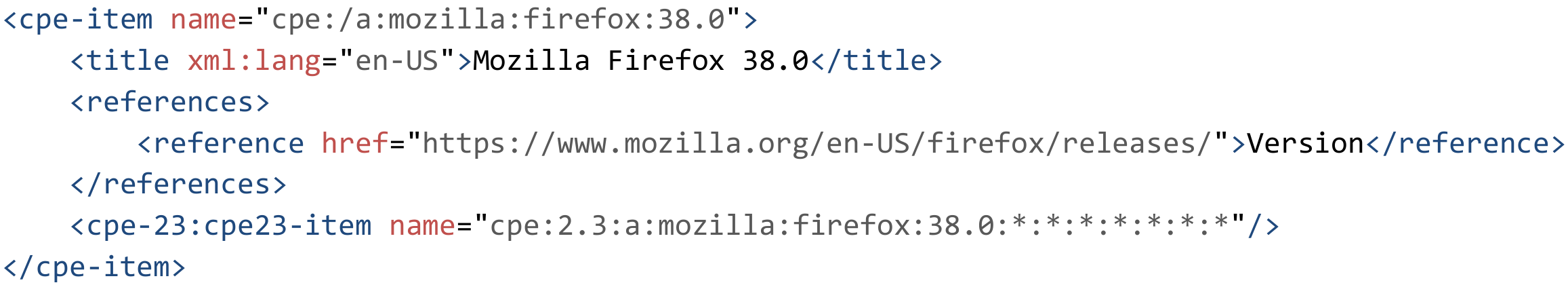}
\caption{CPE entry for Mozilla Firefox 38.0 in the official CPE dictionary}
\label{figure:cpe_entry}
\end{figure*}

The CPE naming scheme is defined by a set of attributes called \emph{Well-Formed CPE Name} (\emph{WFN}). The following attributes are part of this format: \emph{part}, \emph{vendor}, \emph{product}, \emph{version}, \emph{update}, \emph{edition}, \emph{language}, \emph{sw\_edition}, \emph{target\_sw}, \emph{target\_hw}, and other. Listing \ref{listing:wfn} shows the WFN format for the software product \emph{Microsoft Internet Explorer 8}. The value of the attribute \emph{part} indicates that the IT asset is an application. The logical value \emph{NA} means not applicable or not used. \emph{NA} is assigned to attributes that have no meaning for a software product. The logical value \emph{ANY} indicates that there are no restrictions for an attribute.       
\begin{lstlisting}[caption={Possible WFN format for Microsoft Internet Explorer 8},label={listing:wfn},breaklines=true,basicstyle=\footnotesize, language=Python,morekeywords={part, vendor, product, version, update, edition, language, sw_edition, target_sw, target_hw, target_sw, other},breakatwhitespace=true]
part:a, vendor:microsoft, product:internet_explorer, version:8, 
update:NA, edition:ANY, language:ANY, sw_edition:ANY,
target_sw:ANY, target_hw:ANY, other:ANY
\end{lstlisting} 
The CPE specification is used to assign identifiers to assets inside an IT infrastructure. Currently, CPE supports two formats: URI (defined in CPE version 2.2) and formatted string (defined in CPE version 2.3). A URI or formatted string identifier is generated from the WFN of an IT asset. This process is called URI or formatted string binding. Listing \ref{listing:uri_format} and Listing \ref{listing:string_format} show examples for the URI binding and formatted string binding respectively. 
\begin{lstlisting}[caption={WFN of Listing \ref{listing:wfn} bound to a URI},label={listing:uri_format},breaklines=true,basicstyle=\footnotesize, language=Python]
cpe:/a:microsoft:internet_explorer:8:-
\end{lstlisting} 
\begin{lstlisting}[caption={WFN of Listing \ref{listing:wfn} bound to a formatted string},label={listing:string_format},breaklines=true,basicstyle=\footnotesize,language=Python]
cpe:2.3:a:microsoft:internet_explorer:8:-:*:*:*:*:*:*
\end{lstlisting}
Besides the naming specification, the CPE 2.3 stack encompasses the CPE dictionary specification, which defines the structure (e.g., names and metadata) of a repository containing CPE identifiers for classes of IT products. Each entry in the dictionary contains the bound form (URI and/or formatted string) of a product's WFN. The National Vulnerability Database (NVD) hosts and maintains the official CPE dictionary\footnote{\url{https://nvd.nist.gov/products/cpe}}, which is provided in XML format. Figure \ref{figure:cpe_entry} shows a CPE entry of the official dictionary version 2.3. The entry contains the URI (for backward compatibility with CPE version 2.2) and formatted string that identify the application \emph{Mozilla Firefox 38.0}.  
\subsection{Common Vulnerabilities and Exposures (CVE)}
\begin{figure*}[!t]
\centering
\includegraphics[width=4.8in]{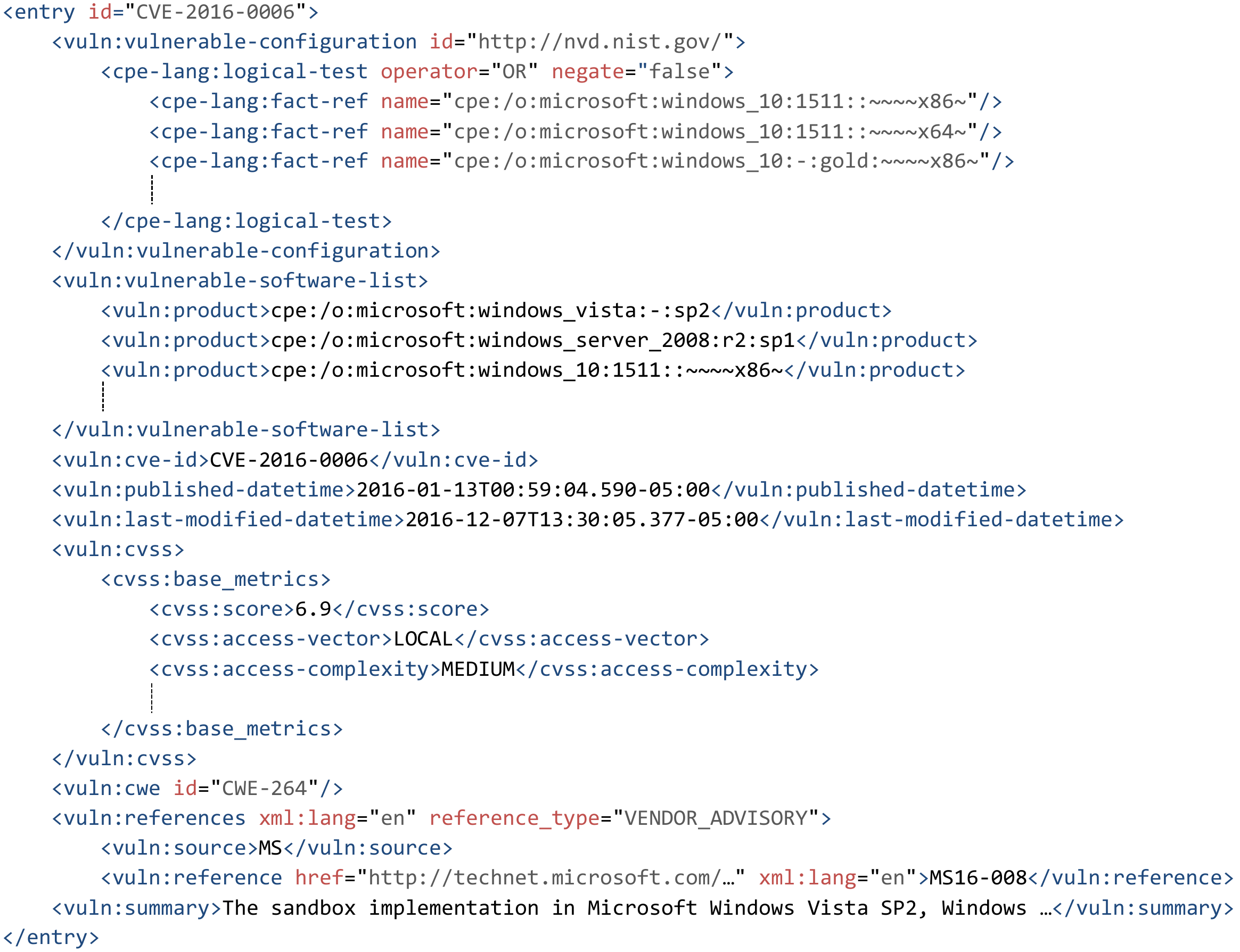}
\caption{Entry of the CVE feed for 2016. The identifier of this vulnerability is CVE-2016-0006}
\label{figure:cve_entry}
\end{figure*}

CVE, which is also part of the SCAP specifications, is a method used to assign identifiers to publicly known vulnerabilities found in IT products and to provide information (e.g., affected products) about the vulnerabilities. Across organizations, anti-virus vendors, and security experts, CVE has become the de facto standard to share information on known vulnerabilities and exposures. Assignment of unique identifiers to vulnerabilities is managed by the MITRE Corporation\footnote{\url{https://cve.mitre.org}}, and the NVD provides documents in XML format (CVE feeds\footnote{\url{https://nvd.nist.gov/vuln/data-feeds}}) that contain the CVE identifiers defined by MITRE along with additional information (e.g., severity, vulnerable configuration, and a list of vulnerable software). Figure \ref{figure:cve_entry} shows a CVE entry obtained from the CVE feed for 2016. Note that some information has been omitted to save space.
\subsection{CVE and CPE in a Vulnerability Management System (VMS)}
One of the goals of a VMS is the detection of vulnerable IT assets (e.g., vulnerable software products) in an IT infrastructure. A VMS also includes other aspects like risk management and remediation. In this paper, we focus on the vulnerability detection capabilities of a VMS. In this context, the CPE dictionary and CVE feeds play an important role. VMSs employing CPE and CVE depend on the compatibility between both datasets. The CPE dictionary is queried by a VMS to find CPE identifiers that match IT products in the infrastructure where the VMS operates. Since the CVE entries of the CVE feeds contain a list of vulnerable software in CPE format (see Figure \ref{figure:cve_entry}), it allows the VMS to find related vulnerabilities for the IT products using their assigned CPE identifier.  

%% file: analysis.tex
\lstset{escapeinside={(*}{*)}}
This section discusses issues that we found in the official CPE dictionary\footnote{\url{https://nvd.nist.gov/cpe.cfm}} and CVE feeds\footnote{\url{https://nvd.nist.gov/download.cfm}} that can lead to incorrect results of VMSs relying on those repositories to find vulnerabilities in software products.   
\subsection{CVE entries without CPE entries}
The CVE feeds (as of February 14th, 2017) contain 895 CVE entries that do not have any CPE identifier. For example, CVE-2016-9748 is a vulnerability for 
the products \emph{IBM Rational DOORS Next Generation 5.0} and \emph{6.0}. This CVE has no CPE identifier, even though the official CPE dictionary has CPE identifiers for these products, as shown in Listing \ref{listing:cpes_ibm_rational_doors}.
\paragraph{Why this is a problem}
VMSs that only use CPE identifiers to find related CVE entries for software products deliver incomplete results. For example, if a vulnerability is found and a CVE entry is created for it without CPE identifiers, a VMS would not generate any alert for the IT assets affected by that vulnerability.
\begin{lstlisting}[caption={CPE identifiers found in the official CPE dictionary (accessed on February 14th, 2017) for IBM Rational DOORS Next Generation versions 5.0 and 6.0.0},label={listing:cpes_ibm_rational_doors},breaklines=true,basicstyle=\footnotesize,language=Python]
cpe:/a:ibm:rational_doors_next_generation:5.0.
cpe:/a:ibm:rational_doors_next_generation:6.0.0
\end{lstlisting}   
\subsection{Software Products without assigned CPE}\label{sw_without_cpe}
There are software products for which no CPE entry has been created in the official CPE dictionary. For example, the product \emph{Microsoft Search Server} does not have an entry in the dictionary. Nonetheless, according to CVE-2008-4032\footnote{\url{https://web.nvd.nist.gov/view/vuln/detail?vulnId=CVE-2008-4032}}, version 2008 of this product is vulnerable, and the CVE entry contains two CPE identifiers for it.
\paragraph{Why this is a problem} Full automation (i.e., without human interaction) of the process of assigning a CPE identifier to a software product can lead to false positives or false negatives. Since there are software products for which no CPE identifier exists in the CPE dictionary, a fully automated system trying to assign an existing CPE to a software product would either assign an incorrect best match or not assign a CPE at all. Consequently, either incorrect CVE entries are found for these software products (false positives), or the actual CVE entries related to those products are not found (false negative).      
\subsection{CPE Dictionary Deprecation Process}
According to the CPE dictionary specification version 2.3 \cite{cpe_2_3_dictionary_specification}, CPE identifiers are deprecated for three reasons: identifier name correction, identifier name removal, or additional information discovery. Up to February 14th, 2017, the CPE dictionary contained 2,614 deprecated entries; all of them are due to name correction. Listing \ref{listing:deprecated_cpe} shows one of the deprecated identifiers found in the official CPE dictionary. In this case, the deprecation was caused by a typo in the word \emph{player}. 
\paragraph{Why this is a problem}
If a VMS assigns a CPE identifier to a software product that is later corrected, then CVEs containing the corrected CPE could not be found. Following the example shown in Listing \ref{listing:deprecated_cpe}, if a system tries to obtain CVE entries that fulfill the condition: vendor equals \emph{adobe} and name equals \emph{flash\_\textbf{playe}\_for\_linux}, it is likely that the query does not return any results, since the typo is probably present only in the CPE dictionary and not in the CVE feeds.
\begin{lstlisting}[caption={Deprecated identifier in the official CPE dictionary},label={listing:deprecated_cpe},breaklines=true,basicstyle=\footnotesize,language=Python]
cpe:/a:adobe:flash_(*\bf playe*)_for_linux:9.0.115.0 (*\bf (deprecated)*)
cpe:/a:adobe:flash_(*\bf player*)_for_linux:9.0.115.0 (*\bf (corrected)*)
\end{lstlisting}
\subsection{NVD Synchronization}
The CVE feeds (as of February 14th, 2017) contain 105,591 CPE entries that do not exist in the CPE dictionary. The analysis showed that this problem is caused by several reasons. One of them was already pointed out in Section \ref{sw_without_cpe}: in the official CPE dictionary, there are vulnerable IT products without CPEs. In addition, there are CPEs which are semantically equal, but nonetheless, their WFNs' attributes contain different values, resulting in different CPEs. Listing \ref{listing:cpe_sematically_equal} shows an example of this case. The CPE found in the CPE dictionary uses the update attribute. On the other hand, the CPE of the CVE feeds concatenates the update (beta1) with the version (1.4.0\_beta1). This clearly demonstrates that synchronization between the official CPE dictionary and the CVE feeds do not take place, though both datasets are hosted in the NVD which is managed by the NIST.
\begin{lstlisting}[caption={CPEs semantically equal but with different WFN attributes. The first CPE is found in the official CPE dictionary and the second one in the CVE feeds.},label={listing:cpe_sematically_equal},breaklines=true,basicstyle=\footnotesize,language=Python]
cpe:/a:digium:asterisk:1.4.0(*\bf:beta1*)
cpe:/a:digium:asterisk:1.4.0(*\bf\_beta1*)
\end{lstlisting}
\paragraph{Why this is a problem} If a VMS assigns a CPE to a software product using the entries of the CPE dictionary only, there is a risk that no CVE is found for that product.   

%% file: method.tex
\begin{figure*}[!t]
\centering
\includegraphics[width=2.5in]{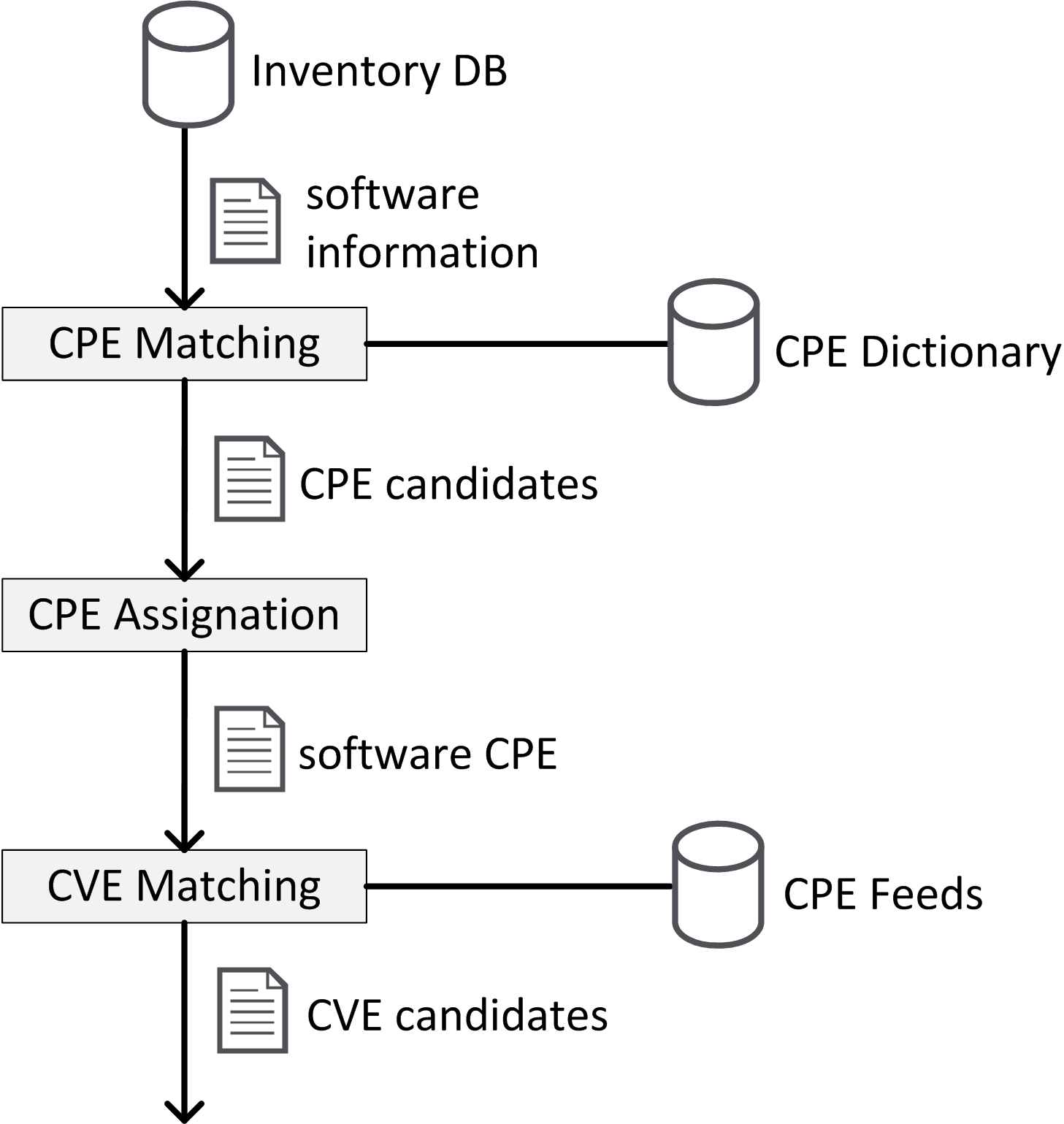}
\caption{Method overview}
\label{figure:overview}
\end{figure*}
The method described in this paper comprises three steps: CPE matching, CPE assigment, and CVE matching. An overview is shown in Figure \ref{figure:overview}.

In the first step, CPE identifiers that could be assigned to a software product (CPE \emph{candidates}) are searched for. The search is based on the product's vendor, name, and version strings (e.g., Microsoft Internet Explorer 8) that is stored in an inventory database. As shown in Section \ref{sec:analysis}, automatically assigning CPEs to software products is problematic. The evaluation in Section \ref{sec:evaluation} supports this finding. Therefore, in the second step, human interaction is required to assign the most suitable CPE to a product. In the third step, CVEs are looked up for a software product based on its assigned CPE. The CPE and CVE matching is carried out using the CPE WFN format. Hence, the URIs of the CPEs are first unbound to WFNs. The described method is the core of an open-source VMS that we will present in Section \ref{sec:evaluation}. In the following sections, the conversion of a URI to a WFN is shortly described, and thereafter, the CPE and CVE matching algorithms are explained.
\subsection{Well-Formed Name (WFN) Conversion}\label{subsection:wfn_conversion}
Before carrying out the steps shown in Figure \ref{figure:overview}, the URIs (CPE identifiers) of both CPE dictionary and CVE feeds are converted into their WFN format. The conversion is based on the CPE Naming Specification Version 2.3 \cite{cpe_2_3_naming_specification}. Listing \ref{listing:uri_binding_wfn} shows the conversion of a URI binding into its WFN format.
\lstset{escapeinside={(*}{*)}}
 \begin{lstlisting}[caption={URI to WFN conversion example},label={listing:uri_binding_wfn},breaklines=true,basicstyle=\footnotesize,language=Python,morekeywords={part, vendor, product, version, update, edition, language, sw_edition, target_sw, target_hw, target_sw, other},breakatwhitespace=true]
CPE URI
cpe:/a:microsoft:internet_explorer:8.*::en~-~~windows~x86~
CPE WFN
part: a, vendor: microsoft, product: internet_explorer, 
version: 8.*, update: ANY, language: en, edition: NA, 
sw_edition: ANY, target_sw: windows, target_hw: x86, other: ANY
\end{lstlisting} 
The idea of this conversion is to compare two CPEs based on their WFN attributes (e.g., vendor) and not using their URIs. This allows to determine which attributes are unequal so that weights can be assigned to the attributes according to their importance. For instance, the attributes vendor, product, and version have higher weights than the attributes target\_sw, target\_hw, or other. In addition, the WFN format allows to perform logical comparisons between two CPEs' versions. For example, the version 8.* is equal to 8.2.7, due to the special character *.

As explained in Section \ref{sec:analysis}, the CPE dictionary can contain incorrect CPE identifiers (e.g., Listing \ref{listing:deprecated_cpe} shows a URI binding with a typo in the word player). This can prevent from finding potential CVE matches for a software product. This can however be overcome with string similarity algorithms such as Levenshtein edit distance \cite{data_matching_christen_peter}. Nonetheless, if we apply this algorithm for the whole URI string, we cannot establish which attributes (e.g., vendor, product, version) of the CPEs being compared are not equal.  
\subsection{CPE Matching}
\begin{figure*}[!t]
\centering
\includegraphics[width=2.5in]{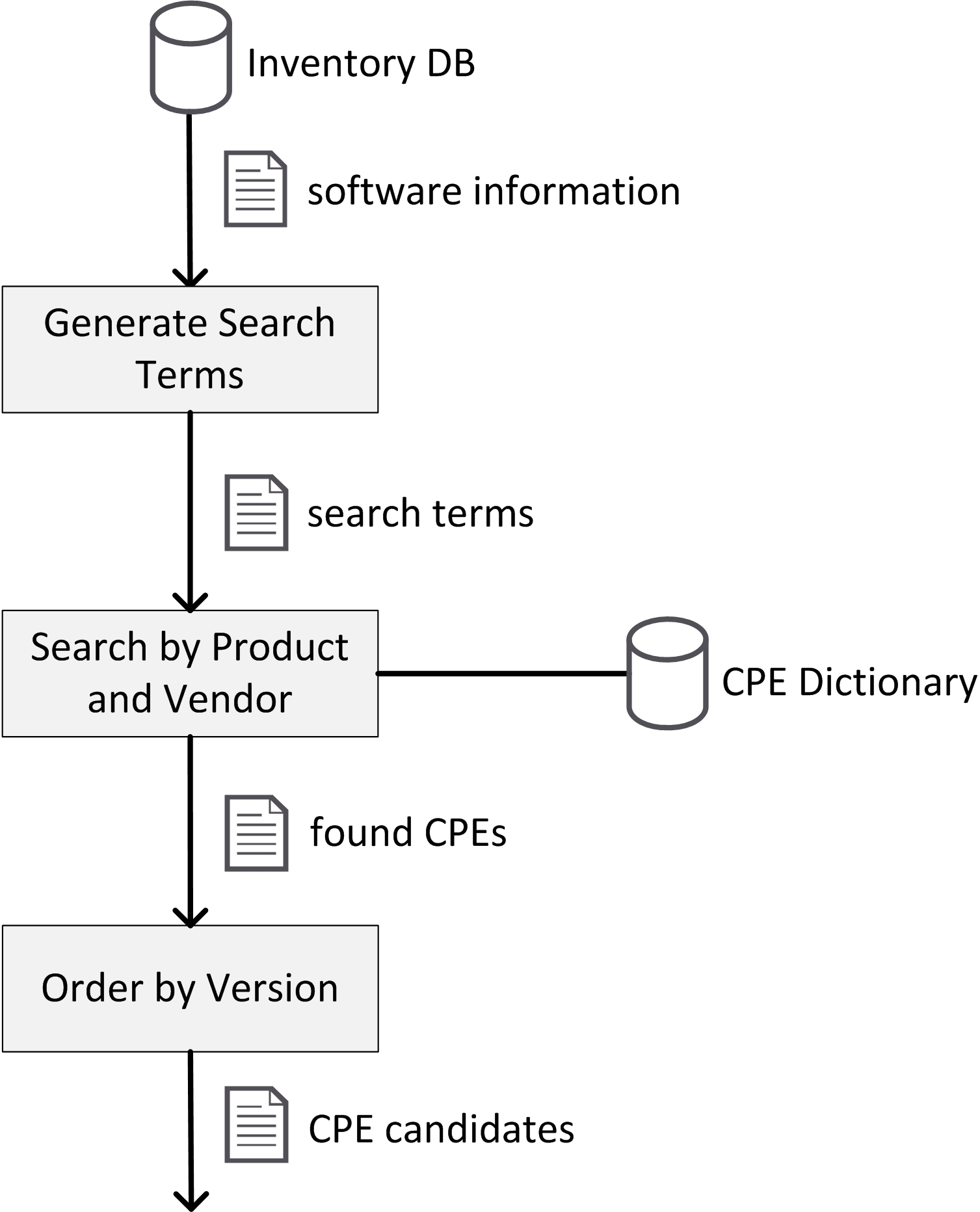}
\caption{CPE matching overview}
\label{figure:cpe_matching_overview}
\end{figure*}

The CPE matching algorithm aids the correct assignment of a CPE to a software product by suggesting CPE candidates found in the CPE dictionary. As depicted in Figure \ref{figure:cpe_matching_overview}, the algorithm receives as input the product's information obtained from an IT assets database (e.g., GLPI\footnote{\url{http://glpi-project.org}}). Then, three steps are performed: \emph{Generate Search Terms}, \emph{Search by Product and Vendor}, and \emph{Sort by Version}.
\paragraph{\textbf{Step 1: Generate Search Terms}}
In this step, using the information provided by the inventory database, search terms to query the CPE dictionary are generated. Listing \ref{listing:software_product_glpi} shows the information obtained from a GLPI database for a Microsoft product. Two groups of search terms are generated: vendor and product search terms. The former are generated with the value of the field ``product'', and the latter with the value of the field ``vendor''. For both groups the following operations are performed: First, the value is split by spaces and converted to lower case. For example, from the string ``Microsoft .NET Framework 4.5.2", the following terms are generated: \emph{microsoft}, \emph{.net}, \emph{framework}, and \emph{4.5.2}. If more than one term is generated, joined terms are formed by concatenating terms with an underscore between them (e.g., \emph{microsoft\_.net\_framework}). Underscores are used in the CPE dictionary to replace spaces. If a vendor term (e.g., \emph{microsoft}) is present in the product search terms, the joined terms of the product are duplicated but removing the vendor term (e.g., \emph{.net\_framework}). During the analysis of the CPE dictionary, we observed that some CPE identifiers include the vendor in the product attribute (e.g., cpe:/a:adobe:\textbf{adobe\_air}:1.0) while other CPE identifiers exclude it (e.g., cpe:/a:adobe:\textbf{air}:15.0.0.293). Therefore, product search terms with and without the vendor are generated. Listing \ref{listing:search_terms} shows the search terms generated with the information of Listing \ref{listing:software_product_glpi}.

The additional search terms increase the search cost but improve the accuracy of the CPE matching algorithm. For instance, to find CPEs similar to \emph{.net\_framework\_4.5.2}, the algorithm should also consider CPEs with a difference of at least six characters (\emph{\_4.5.2}) in the attribute \emph{product} of their WFNs. Doing this would generate inaccurate results, e.g., the string \emph{player} would match the string \emph{joomla}, if the threshold of the edit distance is set to six.    
\begin{lstlisting}[caption={Software product obtained from a GLPI database},label={listing:software_product_glpi},breaklines=true,basicstyle=\footnotesize,language=Python]
(*\textbf{vendor}*): Microsoft Corporation
(*\textbf{product}*):Microsoft .NET Framework 4.5.2
(*\textbf{version}*): 4.5.51209
\end{lstlisting} 
\begin{lstlisting}[caption={Search terms generated with the information shown in Listing \ref{listing:software_product_glpi}},label={listing:search_terms},breaklines=true,basicstyle=\footnotesize,language=Python]
(*\textbf{vendor terms}*): [microsoft_corporation, microsoft, corporation]
(*\textbf{product terms}*):[microsoft_.net_framework_4.5.2, 
microsoft_.net_framework, microsoft_.net, microsoft, 
.net_framework_4.5.2, .net_framework, framework, .net, 4.5.2]
\end{lstlisting} 
\paragraph{\textbf{Step 2: Search by Product and Vendor}}
Using the generated search terms, CPEs for a software product are searched for in the CPE dictionary by comparing the search terms with the CPE WFN attributes \emph{vendor} and \emph{product}. To consider a CPE as match, the following condition must be met: 
\\$(cpe.wfn.vendor \simeq vendor\_search\_term)$ $AND$\\
$(cpe.wfn.product \simeq product\_search\_term)$\\
Note that the symbol similar or equal ($\simeq$) and not the symbol equal ($=$) is employed. As discussed in Section \ref{sec:analysis}, the CPEs can contain typographical errors; therefore, to determine the similarity between a search term and the WFN attribute, the Levenshtein distance algorithm \cite{data_matching_christen_peter} is used. When the Levenshtein distance is less or equal to two, the CPE is considered a match (it can be discarded in the next steps). To establish this threshold, we analyzed the deprecated CPEs in the CPE dictionary. We observed that the CPEs deprecated due to a typo in the attribute product have a Levenshtein distance of one or two. 
\paragraph{\textbf{Step 3: Order by Version}}
Finally, the CPEs found in the previous step are ordered by version. CPEs with similar or equal versions as the  software product version are prioritized. Two versioning schemes are considered: versions defined by years (e.g., 2008) and dotted versions like 1.2.49. The latter are split by the dots.
\subsection{CVE Matching} 
In this section we describe how the CVE matching algorithm searches for CVEs related to a software product.  The algorithm performs three steps: \emph{Search by CVE-CPE Entries}, \emph{Search by CVE Summary}, and \emph{Merge CVE Candidates}.
\paragraph{\textbf{Step 1: Search by CVE-CPE Entries}}
As explained in Section \ref{sec:background}, each entry in the CVE feeds contains a list of software products that are affected by the CVE. The list is written in the form of CPE URIs, which facilitates finding CVEs for a software product using its assigned CPE. To determine whether a CVE is a match for a software product, the product, vendor, and version attributes of the software product's CPE are compared with the product, vendor, and version of each entry of the CVE vulnerable software list. Note that to compare those attributes the URIs of the CVE must be first unbound to the WFN format (see Subsection \ref{subsection:wfn_conversion}).  

Algorithm \ref{algorithm:search_by_cve_cpe_entries} summarizes the process of finding CVEs for a software product in the CVE feeds. The algorithm receives as input the software's CPE and the CVE feeds. It compares the CPEs of each CVE with the software's CPE. To perform the comparison two functions are defined: $similar()$ and $same\_version()$. The function $similar()$ compares two strings (e.g., $similar(microsoft, microsof)$) using the Levenshtein distance algorithm \cite{data_matching_christen_peter} and returns true if and only if the distance is less than three. This allows dealing with possible typographical errors in either the entries of the CPE dictionary or in the entries of the CVE feeds. The function $same\_version()$ compares the version of a software product with the version of a CVE-CPE. The function uses the asterisk symbol to map a version to a group of versions. For example, the version $1.2.*$ is equal to the versions $1.2.3$, $1.2.3.5256$ or any version which starts with $1.2$.

\begin{algorithm*}
\label{algorithm:search_by_cve_cpe_entries}
\KwIn{$cve\_feeds$, $software$}
\KwOut{$cve\_candidates$}
$sw\_product = software.cpe.wfn.product$\\
$sw\_vendor = software.cpe.wfn.vendor$\\
$sw\_version = software.cpe.wfn.version$\\
\For{$cve \in cve\_feeds$}{
	\For{$cve\_cpe \in cve.cpe\_entries$}{
		\If{$similar(sw\_product, cve\_cpe.wfn.product)$ $AND$ $similar(sw\_vendor, cve\_cpe.wfn.vendor)$ $AND$ $same\_version(sw\_version, cve\_cpe.wfn.version)$}{
			$cve\_candidates.add(cve)$
		}
	}
 }
 \caption{Search CVE candidates for a software product by CVE-CPE entries}
\end{algorithm*}
\paragraph{\textbf{Step 2: Search by CVE Summary}}
As pointed out in Section \ref{sec:analysis}, there are CVE entries which do not have CPE entries. To tackle this problem, the summaries of the CVEs (see Figure \ref{figure:cve_entry}) are used to find CVE candidates for a software product. The algorithm compares, using the Levenshtein distance algorithm \cite{data_matching_christen_peter}, all the words of a CVE summary with the product and vendor attributes of the software's CPE WFN. Algorithm \ref{algorithm:search_by_cve_summary} implements this step.
\begin{algorithm*}
\label{algorithm:search_by_cve_summary}
\KwIn{$cves\_without\_cpe\_entries$, $software$}
\KwOut{$cve\_candidates$}
$sw\_product = software.cpe.wfn.product$\\
$sw\_vendor = software.cpe.wfn.vendor$\\
\For{$cve \in cves\_without\_cpe\_entries$}{
	\For{$word \in cve.summary$}{
		\If{$similar(word, sw\_product)$ AND $similar(word, sw\_vendor)$}{
			$cve\_candidates.add(cve)$
		}
	}
 }
\caption{Search CVE candidates for a software product by CVE summary}
\end{algorithm*}
\paragraph{\textbf{Step 3: Merge CVE Candidates}}
In this step, the CVE candidates delivered by the steps ``Search by CVE-CPE entries" and ``Search by CVE Summary" are merged. The merged list is the final result of this phase.

%% file: evaluation.tex
As part of this work, we implemented an open-source VMS software called \emph{Inventory Vulnerability Analysis (IVA)\footnote{\url{https://github.com/fkie-cad/iva}}}, which core is the method described in Section \ref{sec:method}. In this section, we evaluate IVA and show its results when performing fully automated analysis. First, the environment on which IVA was evaluated is described.
\subsection{Environment}
\begin{figure*}[!t]
\centering
\includegraphics[width=3.5in]{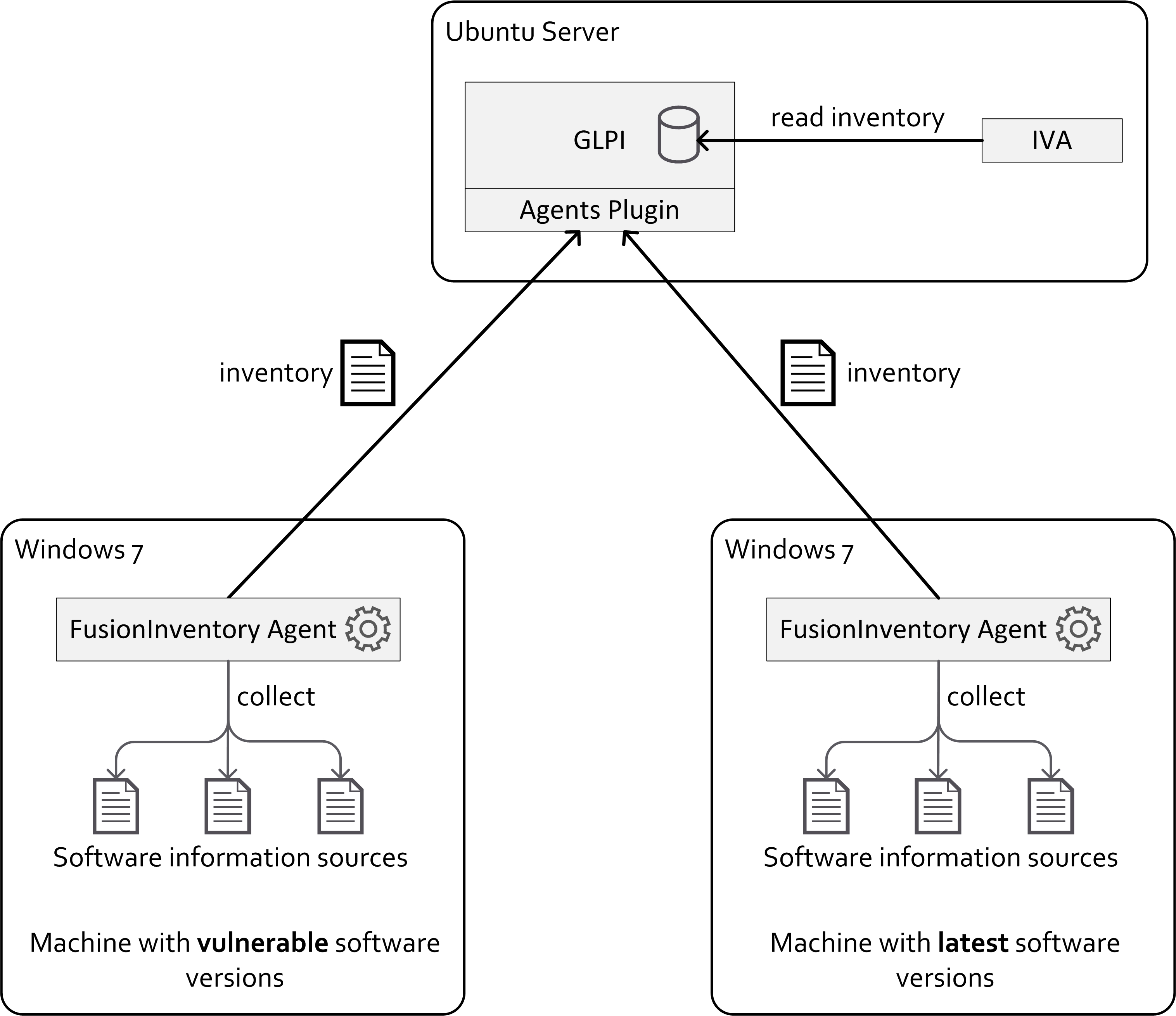}
\caption{Evaluation environment. On the Windows 7 machine on the left side, vulnerable software versions were installed, and on the machine on the right side, the latest versions were installed. GLPI and IVA were set up on an Ubuntu 14.04.4 LTS Server.}
\label{figure:evaluation_environment}
\end{figure*}

To evaluate the proposed method and IVA, we set up an environment (see Figure \ref{figure:evaluation_environment}) that is composed of two Windows 7 virtual machines. On one machine, we installed known vulnerable versions of twelve software products, and on the other machine, the latest versions (at the moment of carrying out the evaluation) of the same products were installed. The list of software products (see Table \ref{table:software_list}) was generated based on the top 50 products by total number of distinct vulnerabilities\footnote{\url{http://www.cvedetails.com/top-50-products.php}} (the list was accessed on March 24rd 2017). From the 50 products, only those compatible with Windows 7 was considered. To collect the information of the software installed on the Windows 7 machines, FusionInventory Agent\footnote{\url{http://fusioninventory.org}} was employed. This tool gathers information from different sources (e.g., the registry) and sends it to an inventory database (in this case, GLPI) in XML format. The software information is then read by IVA from the GLPI database.
\begin{table}
\small
\begin{center}
\begin{tabular}{|l|l|l|l|l|}
\hline
& \textbf{Vendor} & \textbf{Product} & \textbf{Vulnerable}  & \textbf{Latest} \\
\hline
P1 & Mozilla & Firefox & 48.0.2  & 52.0  \\
\hline
P2 & Adobe & Flash Player & 23.0.0.207  & 25.0.0.127  \\
\hline
P3 & Microsoft & Internet Explorer & 8.0.7601.17514  &  11.0.9600.17843 \\
\hline
P4 & Mozilla & Thunderbird & 38.6.0 & 45.8.0 \\
\hline
P5 & Mozilla & SeaMonkey  & 2.35 & 2.46 \\
\hline
P6 & Adobe & Adobe Reader & 11.0.17 & 15.023.20070 \\
\hline
P7 & Oracle & Java & 8.0.1120.15 & 8.0.1210.13 \\
\hline
P8 & Oracle & Java SE Development Kit & 8.0.1120.15 & 8.0.1210.13 \\
\hline
P9 & Wireshark & Wireshark & 2.0.0 & 2.2.5 \\
\hline
P10 & Apple Inc. & iTunes & 12.1.3.6 & 12.5.5.5 \\
\hline
P11 & Oracle & MySQL Server & 5.7.15 & 5.7.17 \\
\hline
P12 & Adobe & Adobe AIR & 20.0.0.260 & 25.0.0.134 \\
\hline
\end{tabular}
\end{center}
\caption{List of software products used for the evaluation of IVA. The fourth column shows the vulnerable versions of the products, and the fifth column shows the latest versions (as of March 24th 2017).}
\label{table:software_list}
\end{table}

\begin{table}
\small
\begin{center}
\begin{tabular}{|l|l|l|}
\hline
\textbf{Vendor} & \textbf{Product} & \textbf{Version} \\
\hline
Adobe Systems & Adobe Flash Player & 23.0.0.207\\ 
Incorporated  & 23 NPAPI & \\
\hline
Microsoft Corporation & Internet Explorer & 8.0.7601.17514\\
\hline
Oracle Corporation & Java SE Development & 8.0.1120.15\\ 
				   & Kit 8 Update 112 & \\
\hline
The Wireshark & Wireshark 2.0.0 (32-bit) & 2.0.0\\
developer community, & & \\
https://www.wireshark.org & & \\ 
\hline
\end{tabular}
\end{center}
\caption{Examples of the software information collected by FusionInventory Agent from the Windows 7 machines}
\label{table:examples_fusioninventory_agent}
\end{table}
\subsection{Fully Automated Vulnerability Analysis}
In this context, fully automated vulnerability analysis refers to the capability of a system to assign a CPE identifier to a software product and then scan for related vulnerabilities without human interaction. In this section, we analyze the results of IVA when assuming that a user assigns the first (best) CPE candidate to a software product. Since IVA was not designed to do this assignment automatically, we simulate this behaviour. Moreover, we do not only analyze the accuracy of IVA to assign a CPE for a software product, but also its capability to find CVEs related to the assigned CPE.    

Table \ref{table:results_vulnerable_sw} and Table \ref{table:results_latest_versions_sw} show the results for the vulnerable and latest versions of the software products installed on the Windows 7 machines. The first column shows the product label according to Table \ref{table:software_list} (e.g., P1 is the label for Mozilla Firefox). The second column represents the best CPE candidate found by IVA for a given product. In this case, this is the CPE that would be automatically assigned to the product. The third column indicates whether the CPE completely matches the product information collected by FusionInventory Agent. Note that the collected information differs from the one shown in Table \ref{table:software_list}. Table \ref{table:examples_fusioninventory_agent} shows examples of the actual information collected by FusionInventory Agent. The fourth column shows the number of CVEs that IVA found related to an assigned CPE. The fifth column shows the number of positives, i.e., CVEs that have CPE entries exactly matching the assigned CPE. The last column represents the number of false positives, CVEs with CPE entries not exactly matching the assigned CPE. For example, if a CVE has CPE entries where only the main version matches the version of the assigned CPE (e.g., $2.35 \neq 2.32$), then the CVE is considered as a false positive.

\begin{table}
\small
\begin{center}
\begin{tabular}{|l|l|c|c|c|c|}
\hline
\multirow{2}{*}{} &\multicolumn{2}{|c|}{\textbf{CPE Matching}}&\multicolumn{3}{|c|}{\textbf{CVE Matching}} \\
\cline{2-6}
 &\textbf{Assigned CPE}&\textbf{Correct?}&\textbf{CVEs}&\textbf{Correct}&\textbf{Incorrect}\\
\hline
P1 & cpe:/a:mozilla:firefox:48.0.2 & Yes & 19 & 18 & 1\\
\hline
P2 & cpe:/a:adobe:flash\_player:23.0.0.207 & Yes & 39 & 17 & 22\\
\hline
P3 & cpe:/a:microsoft:internet\_explorer: & Yes & 349 & 340 & 9\\
   & 						 8.0.7601.17514 &&&&\\
\hline
P4	& cpe:/a:mozilla:thunderbird:38.6.0	& Yes	& 27 & 9 &18\\
\hline
P5	& cpe:/a:mozilla:seamonkey:2.35	& Yes & 475	& 1 & 474\\
\hline
P6	& cpe:/a:adobe:adobe\_reader:11.0.17 & No & 237 & 75 & 162\\
	& ::$\sim\sim\sim$android$\sim\sim$	 &	  &		&	 &	  \\
\hline
P7	& cpe:/a:oracle:jre:8.0.1120.15 & Yes & 0 & 0 & 0\\
\hline
P8  & cpe:/a:oracle:jdk:8.0.1120.15	& Yes &0 & 0 & 0 \\
\hline
P9  & cpe:/a:wireshark:wireshark:2.0.0 & Yes & 85 & 84 & 1 \\
\hline
P10 & cpe:/a:apple:itunes:12.1.3.6 1 & Yes & 146 & 3 & 143 \\
\hline
P11 & cpe:/a:oracle:jserver:5.7.15 & No & - & - & - \\
\hline
P12 & cpe:/a:adobe:adobe\_air:20.0.0.260	& Yes &64 &23 &41\\
\hline
\end{tabular}
\end{center}
\caption{Evaluation results for vulnerable versions of twelve software products installed on Windows 7}
\label{table:results_vulnerable_sw}
\end{table}

\begin{table}
\small
\begin{center}
\begin{tabular}{|l|l|c|c|c|c|}
\hline
\multirow{2}{*}{} &\multicolumn{2}{|c|}{\textbf{CPE Matching}}&\multicolumn{3}{|c|}{\textbf{CVE Matching}} \\
\cline{2-6}
 &\textbf{Assigned CPE}&\textbf{Correct?}&\textbf{CVEs}&\textbf{Correct}&\textbf{False}\\
\hline
P1	& cpe:/a:mozilla:firefox:52.0 & Yes & 0 & 0 & 0\\
\hline
P2	& cpe:/a:adobe:flash\_player:25.0.0.127 & Yes & 0 & 0 & 0\\
\hline
P3	& cpe:/a:microsoft:internet\_explorer:  & Yes & 525	& 523 & 2\\
	& 						 11.0.9600.17843:- &  & 	&  & \\
\hline
P4	& cpe:/a:mozilla:thunderbird:45.8.0	& Yes & 0 & 0 & 0\\
\hline
P5	& cpe:/a:mozilla:seamonkey:2.46	& Yes & 475	& 0 & 475\\
\hline
P6	& cpe:/a:adobe:acrobat\_reader\_dc: & No & 259 & 0 & 259\\
	& 15.023.20070::$\sim\sim$classic$\sim\sim\sim$	 &	  &		&	 &	  \\
\hline
P7	& cpe:/a:oracle:jre:8.0.1210.13 & Yes & 0 & 0 & 0\\
\hline
P8  & cpe:/a:oracle:jdk:8.0.1210.13	& Yes &0 & 0 & 0 \\
\hline
P9  & cpe:/a:wireshark:wireshark:2.2.5 & Yes & 85 & 0 & 85 \\
\hline
P10 & cpe:/a:apple:itunes:12.5.5.5 & Yes & 146 & 0 & 146 \\
\hline
P11 & cpe:/a:oracle:jserver:5.7.17 & No & - & - & - \\
\hline
P12 & cpe:/a:adobe:adobe\_air:25.0.0.134 & Yes & 0 & 0 & 0\\
\hline
\end{tabular}
\end{center}
\caption{Evaluation results for the latest versions of twelve products installed on Windows 7}
\label{table:results_latest_versions_sw}
\end{table}

\subsubsection{CPE Matching}
 
\begin{figure*}[!t]
\centering
\includegraphics[width=3.2in]{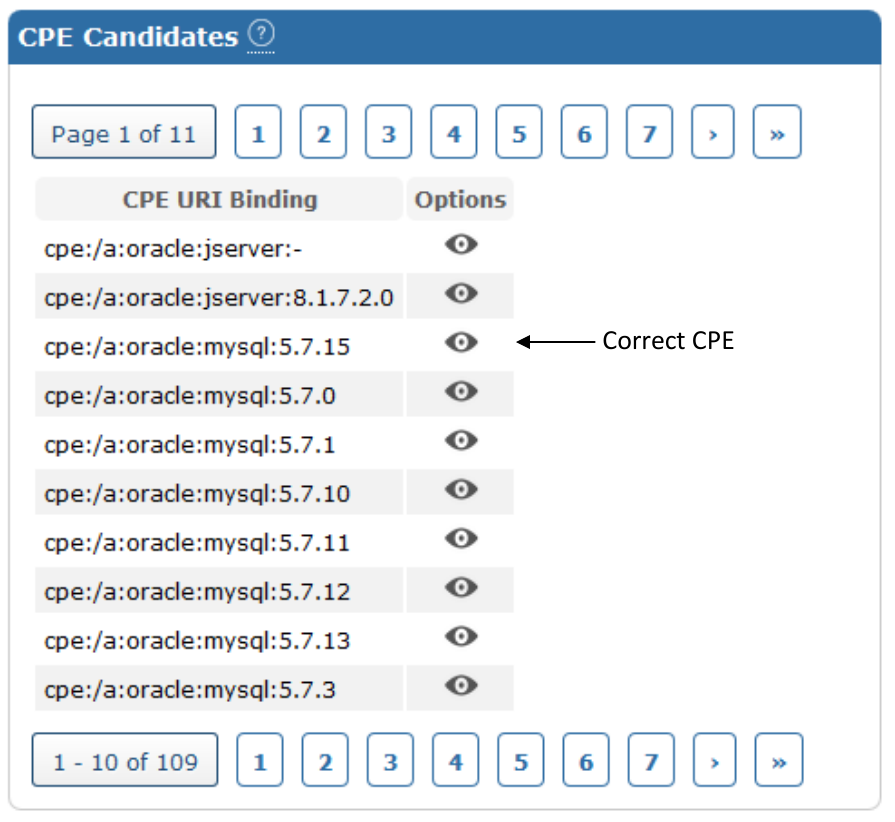}
\caption{IVA screenshot: Top ten CPE candidates for Oracle MySQL Server 5.7.15. The correct CPE is in the third position}
\label{figure:cpe_candidates}
\end{figure*}

As can be seen in Table \ref{table:results_vulnerable_sw}, IVA would be able to automatically assign the correct CPEs to ten products; however, to the products P6 and P11 incorrect CPEs would be assigned. In the former case, the target software (android) is incorrect, and in the latter case, the product (jserver) is incorrect. As depicted in Figure \ref{figure:cpe_candidates}, the correct CPE (cpe:/a:oracle:mysql:5.7.15) for P11 is suggested by IVA in the third position of the candidates list. Likewise, for the latest versions of the software products, ten out of twelve products were assigned correctly. For P6, the software edition (classic) is not the correct one. In this case, the edition should be ``Continuous''. 

The results show that the method described in Section \ref{sec:method} fails to assign correct CPEs to some software products if the assignment is performed automatically. One problem is that the CPE dictionary does not contain CPEs for all products, as analyzed in Section \ref{sec:analysis}. Furthermore, it is a challenging task to find the correct CPEs using the software information collected from a computer. In some cases the information of a software product significantly differs from its CPE. For example, in the case of Wireshark, the system must match the string ``Wireshark" with the string ``The Wireshark developer community, https://www.wireshark.org''. For Java, the system must match the string ``jdk'' with the string ``Java SE Development Kit 8 Update 112''.   

\subsubsection{CVE Matching}
In this section, we evaluate the ability of IVA to find CVEs related to a software product using its automatically assigned CPE. We call a CVE \emph{correct} if the software is actually affected by it, and \emph{incorrect} otherwise. 

For a vulnerable version of a software product, IVA should find at least one correct CVE. Nevertheless, as shown in Table \ref{table:results_vulnerable_sw}, this is not the case for all products. Due to inconsistencies between the version collected by FusionInventory Agent, the version schemes defined by the product's vendor, and the version used in the CPE dictionary and CVE feeds, no CVE was found for the Java products (P7 and P8). As of version 5, a new versioning system was introduced for Java products\footnote{\url{https://en.wikipedia.org/wiki/Java_version_history\#Versioning_change}}. For example, 5 refers to the product version and 1.5 is the developer version. However in the information gathered by the inventory agent from the Windows machine neither the product version nor the developer version is used. For instance, for Java 8 update 112, the version 8.0.1120.15 was provided. The first candidate suggested by IVA keeps the version collected by the inventory agent, but in the CVE feeds all the CPEs for Java products use the developer version (e.g., 1.8.0). Therefore, no CVEs are found for JDK and JRE when employing the version collected by the agent.         

\begin{figure*}[!t]
\centering
\includegraphics[width=3.2in]{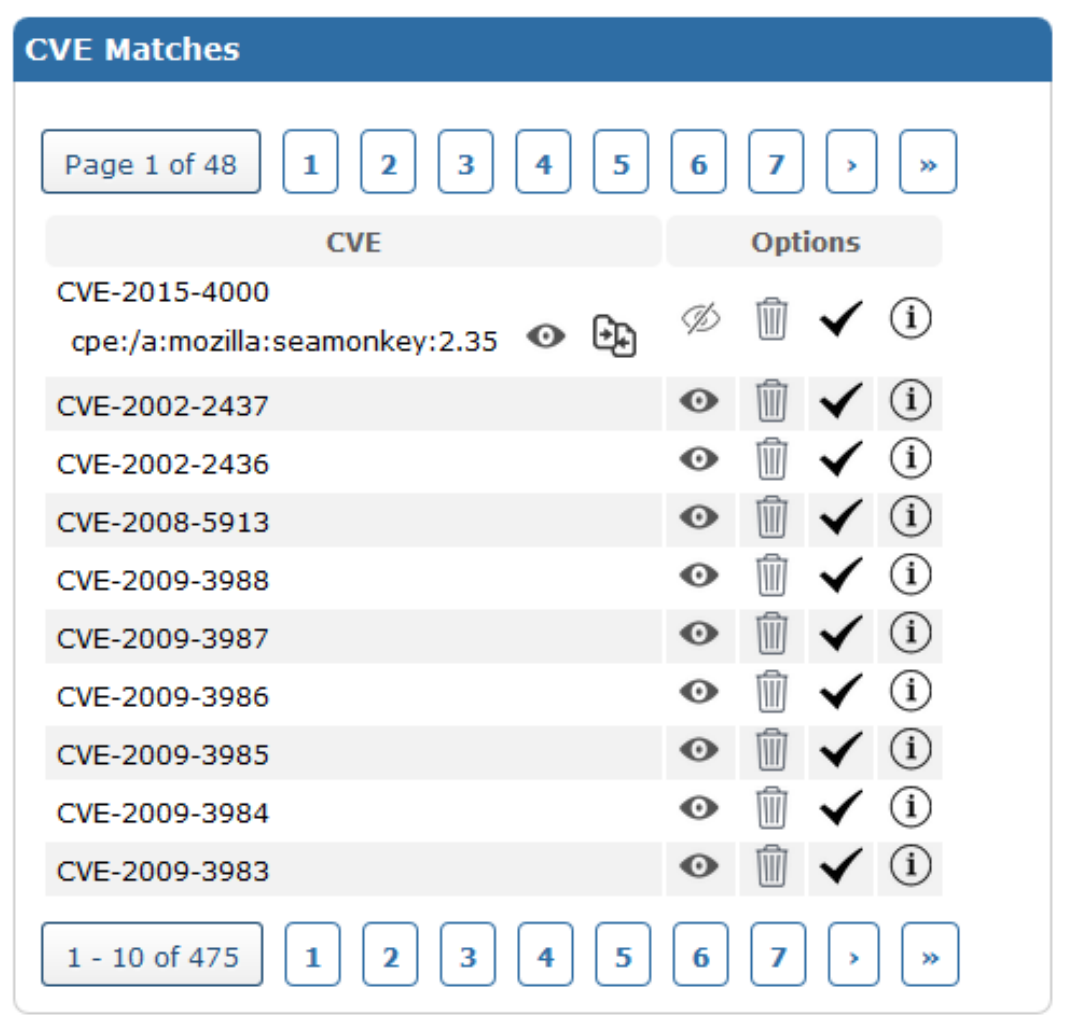}
\caption{IVA screenshot: Top ten CVE matches for Mozilla SeaMonkey 2.35}
\label{figure:cve_candidates}
\end{figure*}

We can also observe in Table \ref{table:results_vulnerable_sw} that some products (e.g., P5) have a high number of incorrect CVEs. This is the case especially when the main version of different product releases are the same (e.g. 2.35 and 2.56), since IVA only considers the main version to suggest CVEs. Nevertheless, IVA prioritizes CVE CPEs with equal version as the version of the software product. Figure \ref{figure:cve_candidates} shows the CVE matches found by IVA for Mozilla SeaMonkey 2.35. As one can see, the first CVE match has a CPE with the same version as the software product. Also note that only the CPEs that are related to the software product are shown. 

\begin{figure*}[!t]
\centering
\includegraphics[width=4.2in]{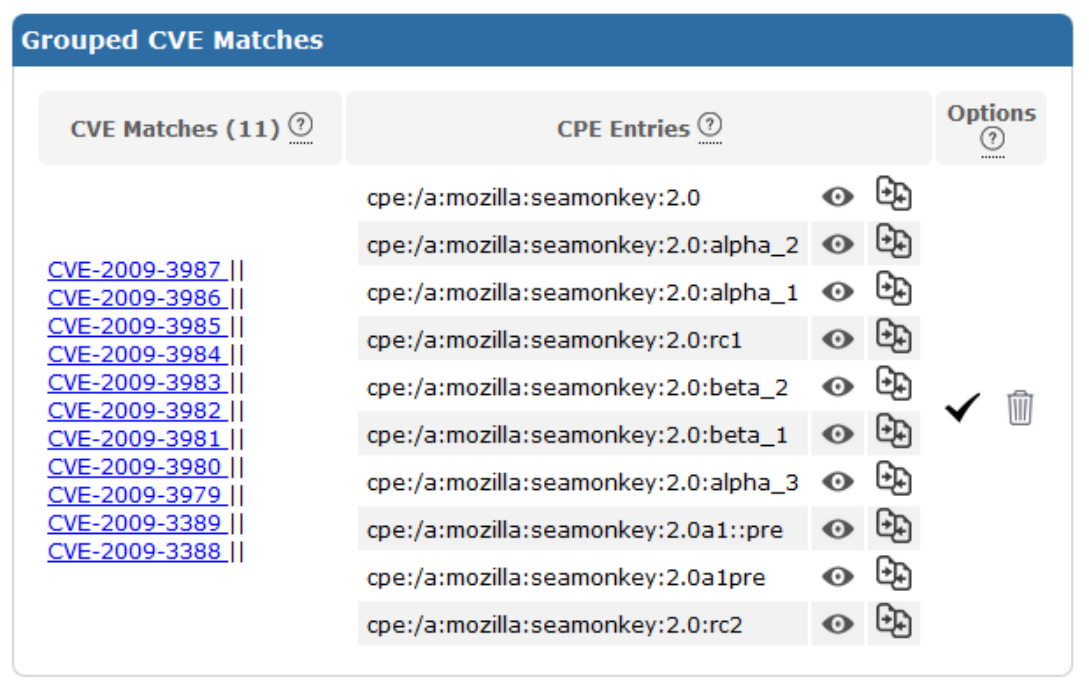}
\caption{IVA screenshot: Grouped CVE matches for Mozilla SeaMonkey 2.35}
\label{figure:grouped_cve_matches}
\end{figure*}

When CVE matches for a software product are found, IVA does not automatically generate alerts. The user must confirm that a CVE is indeed a vulnerability for the software product. To facilitate this task, IVA can group CVEs with equal CPEs. Thus, the user can confirm or discard several CVEs at the same time. Figure \ref{figure:grouped_cve_matches} shows an example of grouped CVEs for Mozilla SeaMonkey 2.35. In this case, the user can discard all CVEs since they do not have any CPE that matches the version (2.35) of the product. 

To the products P6 and P11 incorrect CPEs were assigned. Nonetheless, since IVA compares only the vendor, product, and version, for P6 75 correct CVEs were found. This is not the case for P11, since ``jserver" does not correspond to the correct product name ``mysql".    

For the latest versions of the software products (see Table \ref{table:results_latest_versions_sw}) we analyze the following cases. For the products P1, P2, P4, and P5, no CVE was found, which is the result that is expected for the latest versions of software products. For P5, P6, P9, P10 CVEs were found, but none of them are correct. As explained before, when determining whether two CPEs are equal or not, apart from the vendor and product name, IVA only considers the main version (e.g., 2.35 $=$ 2.45). The results for the Java products (P7 and P8) seem to be correct; however, as explained before, IVA employed the version collected by the inventory agent (e.g., 8.0.1120.15), which format does not correspond to the one used in the CVE feeds for Java products (1.8.0). 

Even though the latest version of Internet Explorer (version 11) was installed, 523 CVEs were found for P3. In these CVEs, only the main version of Internet Explorer 11 (e.g., cpe:/a:microsoft:internet\_explorer:11) is mentioned. The CVEs description indicates however which security updates must be installed to patch the vulnerabilities. Therefore, in this case, IVA just provides a hint to the user that the software product can be vulnerable. The user needs to verify whether the required security updates have been installed.  

%% file: related_work.tex
Takahashi et al. describe a system and architecture to monitor software vulnerabilities in a company's network \cite{takahashi}. Similar to the method explained in this paper, the CPE dictionary is employed to look up CPE identifiers to be assigned to software products based on the information collected by agents and network scanners. In addition, the CVE feeds are used to find vulnerabilities in software products using the assigned CPEs. The proposed technique aims at carrying out this task without human interaction. The aspects discussed in Section \ref{sec:analysis} and Section \ref{sec:evaluation} are not considered. Among others, the inconsistencies in the official CPE dictionary and CVE feeds could lead to a poor efficacy, risking the security of an IT infrastructure. When searching for CVEs, it is assumed that the correct CPE was assigned. Furthermore, the employed CPE and CVE matching algorithms are not explained. 

Torkura and Meinel propose an approach to improve vulnerability assessment for IT assets deployed on cloud environments (e.g., the Elastic Computing Cloud) \cite{torkura_meinel}. They implemented a system using OpenVAS\footnote{\url{http://www.openvas.org/index.de.html}} as a vulnerability scanner, and NVD and Amazon Web Service (AWS) security advisories as vulnerability information sources. Other sources of information are employed, since, based on previous studies, the authors are aware of the inaccuracy of the NVD datasets. However, solutions to deal with the NVD drawbacks are not proposed. Based on this approach, a system called CAVAS is implemented, and its ability to correctly detect cloud components (e.g., Amazon Linux) is evaluated. Nonetheless, an evaluation on the capability of CAVAS to correctly find vulnerabilities in the detected components is not provided.

Other works have already pointed out the disadvantages when using the NVD datasets for vulnerability management. Zhang et al. presented a study in which it is observed that the NVD is not adequate to find software vulnerabilities due to missing software information, different version schemas, vendors' bad practices in vulnerability release time, and data errors \cite{zhang}. By means of an experiment, Nguyen and Massacci demonstrated that a significant number of CVEs for Google Chrome erroneously refer to some other versions \cite{nguyen_massacci}. Fitzgerald and Foley show that inconsistencies in SCAP can cause name ambiguities in the CPE dictionary. They explain that the CPE naming specification \cite{cpe_2_3_naming_specification} does not offer a mechanism to define relationships between CPEs, and thus, ambiguities are likely to exist \cite{fitzgerald_foley}.

%% file: conclusion.tex
In this paper, we analyzed the official CPE dictionary and the CVE feeds. Four major issues were observed: CVE entries without CPE references, software products without assigned CPEs, typographical errors, and a lack of synchronization between both datasets. To cope with these problems, we proposed a novel method comprised of three steps: CPE matching, CPE assignment, and CVE matching. The method is based on string similarity algorithms (e.g., edit distance), and moreover, instead of using the URI of a CPE, its WFN format is employed, which allows to determine the similarity of each attribute (e.g., vendor, product, version) separately between two compared CPEs. We also presented IVA (\emph{Inventory Vulnerability Analysis)}, a Vulnerability Management System that implements the proposed method. IVA is available as open-source software on GitHub\footnote{\url{https://github.com/fkie-cad/iva}}}.

Our evaluation (employing IVA) showed that a fully automated assignment of a CPE identifier to a software product is impractical and error prone. IVA therefore proposes a prioritized list of CPE candidates and lets the user select (and, if necessary, change) the most suitable CPE.

During the evaluation other obstacles that can hinder the process of vulnerability analysis were encountered: name ambiguities in the CPE dictionary, versioning format inconsistencies, and significant differences between the information collected by the inventory agent and the information in the CPE and CVE repositories for a software product. The erroneous assignment of a CPE identifier to a software product can prevent from finding CVEs related to that product, thus increasing the risk of a successful cyberattack. Therefore, when using the CPE dictionary and CVE feeds for vulnerability scanning, the issues discussed in this paper should be considered. Moreover, the NIST, which is responsible for maintaining these repositories, should define a mechanism (e.g., synchronization between the datasets when adding, modifying or deleting a CPE) to overcome these issues.